\documentclass[]{spie}

\usepackage{amsmath,amsfonts,amssymb}
\usepackage{graphicx}
\usepackage[colorlinks=true, allcolors=blue]{hyperref}
\usepackage{astro_bib_macro}

\title{SPIE Proceedings: Characterization and on-sky testing of photonic, AWG-based astronomical spectrographs}

\author[a]{Jerónimo Calderón-Gómez*}
\author[a]{Frantz Martinache}
\author[a]{Nick Cvetojevic}
\author[a]{Marc-Antoine Martinod}
\author[a]{Vincent Foriel}
\author[b]{Nemanja Jovanovic}
\author[c]{Pradip Gatkine}
\author[b]{Gregory Sercel}
\author[d]{Alexis Carlotti}
\author[d]{Marc Ferlet}
\author[e]{Matteo Pasinetti}
\author[e]{Angelie Alagao}
\author[e]{Arnaud Striffling}
\author[d]{Guillermo Martin}
\author[a]{David Mary}
\author[a]{Roxanne Ligi}
\author[a]{Sylvie Robbe-Dubois}

\affil[a]{Université Côte d'Azur, Observatoire de la Côte d'Azur, CNRS, Laboratoire Lagrange, Bd de l'Observatoire, CS 34229, 06304 Nice cedex 4, France}
\affil[b]{Department of Astronomy, California Institute of Technology, 1200 E. California Blvd., Pasadena, CA 91125, USA}
\affil[c]{Department of Physics \& Astronomy, University of California, Los Angeles (UCLA), 475 Portola Plaza, Los Angeles 90095, USA}
\affil[d]{Université Grenoble Alpes, CNRS, IPAG, 38000 Grenoble, France}
\affil[e]{Aix Marseille Université, CNRS, LAM (Laboratoire d'Astrophysique de Marseille) UMR 7326, F-13388 Marseille, France}

\authorinfo{*: Corresponding author: \href{mailto:jcalderon@oca.eu}{jcalderon@oca.eu}}
\begin{document} 
\maketitle

\begin{abstract}
Astrophotonics is a field that intends to meet the needs of next-generation instruments at a small footprint, low cost, and high stability, compared to bulk-optics-based alternatives. Much development effort is driven by the stringent requirements of direct detection and characterization of exoplanets.
Our team works in characterizing Arrayed-Waveguide Grating (AWG) chips for photonics-based, high-resolution, near infrared spectro-interferometry. AWG spectrographs allow to test the feasibility of photonic spectro-interferometers for exoplanet characterization, another step towards fully photonic instruments for astronomy.
We present the current status of our AWG characterization and a preliminary on-sky qualification campaign at the PAPYRUS AO system. We present the CoLiBRIS-AWG spectrograph prototype built for on-sky testing, and preliminary results using our high-resolution (H band, $R\sim 18000$) AWG for observations of Arcturus ($\alpha$ Boötes) and Betelgeuse ($\alpha$ Orionis). This work contributes to assessing the capabilities of photonic spectroscopy for the development of future compact instruments.

\end{abstract}


\section{INTRODUCTION}
\label{sec:intro}

The development of fiber optics and waveguiding technologies for telecommunications has heavily benefited the newly conceived field of astrophotonics. The use of fiber optics has become more common in astronomical observatories, mostly for remapping the light of a source on the focal plane of the telescope into a spectrograph or other downstream instruments. Besides this, photonic fibers in astronomy allow to efficiently transport light long distances and reconfigure it at the point of detection at will.

A more recent development towards the design and production of compact ``instruments on a chip", is the use of photonic integrated chips (PICs) for astronomy: microscopic waveguides can be manufactured within glass substrates in order to accurately and reliably manipulate light signals. These chips are light-weight, non-expensive, and easy to mass-produce once the prototypes have been proven to work. Their use has quickly become extensive in the recent decades for the telecommunications industry, providing a head start in the reliability and reproducibility of these technologies for astronomical research. 

Relevant capabilities of PICs useful for astronomy include spectral dispersion, optical filtering, phase or amplitude modulation, light generation, frequency shifting, light detection, and many more. These photonic circuits can involve passive or active elements that allow precise manipulation of light without moving parts or demanding maintenance needs, common in their bulk optics counterparts. Technologies like Photonic Lanterns, Laser Frequency Combs, photonic beam combiner chips for interferometry, and others already have seen satisfactory applications in observatories around the world\cite{Jovanovic2023}.

In the case of photonic spectroscopy, one of the key goals of the field is the development of an Integrated Photonic Spectrograph (IPS)\cite{Jovanovic2023}, and to achieve them, the photonic dispersive elements had to be developed and tested. In this work we approach the characterization of one of such components: the Arrayed Waveguide Grating (AWG). A chip architecture that achieves the dispersion of light through a principle analogous to that of an échelle grating, but covering a surface of only a few cm$^2$ while easily achieving high spectral resolution ($R \gtrsim 10000$).

In previous years, photonic dispersive technologies like the AWG have captured the attention of the astrophotonics community, leading to the development of the first AWG-based spectrographs used in labs and for astronomical observations\cite{Cvetojevic2012,Gatkine2017,Gatkine2021}. This work was crucial to demonstrate the readiness and availability of the technology for astronomical applications but more extensive on-sky testing with an instrument involving AWGs has not been reported yet. This justifies the need for a dedicated and versatile platform for testing these devices at the focus of a telescope with Adaptive Optics (AO) capabilities through the observation of well known stellar targets with relevant spectral features that can showcase the potential of AWGs for astronomy.

In this work we present the AWG chips available in our laboratory used for the design and construction of a first prototype, the in-lab characterization procedures followed, and the preliminary results of an on-sky qualification campaign involving AWGs at the heart of an astronomical spectrograph.

\section{Arrayed-waveguide gratings (AWG)}

\begin{figure}[t]
    \centering
    \includegraphics[width=1\linewidth]{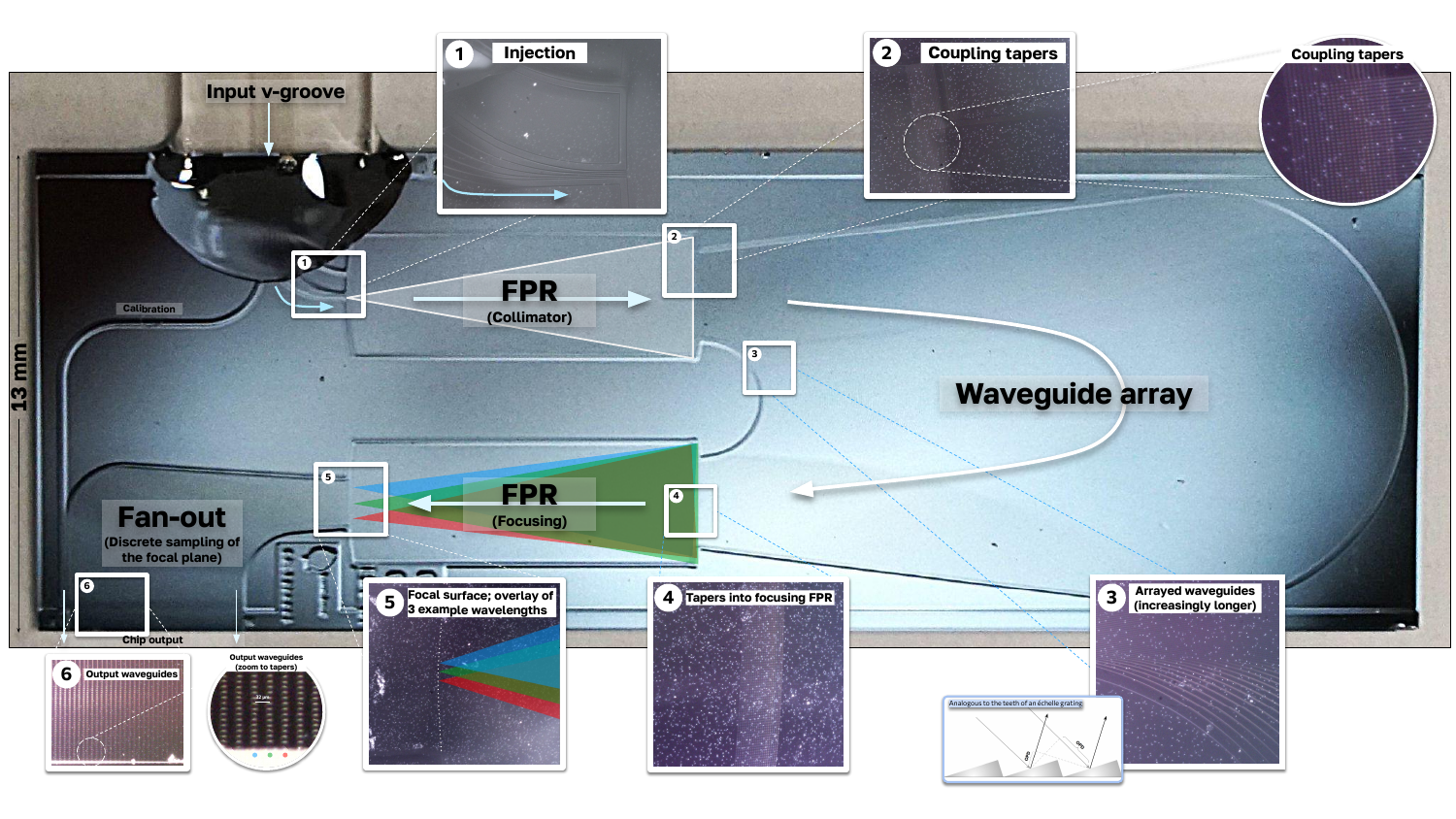}
    \caption{Detailed view of the Emerald AWG chip and its main components. Arrows show the path followed by the light inside the chip. Light enters the chip from the input v-groove at the top-left part of the image.}
    \label{fig:AWG_Emerald_Detailed}
\end{figure}

AWGs are among the technologies developed for what the telecommunications industry calls \textit{wavelength division multiplexing} (WDM). AWGs are photonic dispersers that work by guiding light from their input through tens or hundreds of single-mode waveguides with path-length differences carefully designed to generate a spectrally dispersed signal at its output. The concept was first reported by Smit (1988) under the name of Phased Arrays (PHASARs)\cite{Smit1988}, they were also called Waveguide Grating Routers (WGRs), and soon after with pioneering work by Takahashi (1990) the term Arrayed Waveguide Grating was coined\cite{Takahashi1990}. 

In the telecommunications industry, this device is useful for its multiplexing/demultiplexing capabilities: multiple signals can be encoded into different spectral channels and combined onto the same beam with an AWG at the transmitter, and with a reversed setup the signal can be dispersed again and recorded at the receiver. This is advantageous to optimize the amount of fibers needed to send multiple signals simultaneously, which resulted in extensive applications in the industry\cite{Leijtens2006}.

A natural question that follows this initial application is if a reliable, efficient, and compact astronomical spectrograph involving AWGs is feasible. State-of-the-art, high-resolution spectrographs are known to be bulky systems. This is due to what is known as the telescope-spectrograph size relation: for seeing limited observations, spectrographs increase in size proportional to the diameter of the telescope aperture. If observations are instead performed in conditions close to the diffraction limit, the size relation is broken and the focused PSF of the telescope has a size comparable to the input fibers of photonic chips. Additionally, in conventional grating-based spectrographs the spectral resolution is proportional to the number of illuminated grooves on the grating, which ends up increasing the size of the components involved in the instrument. It becomes then a matter of efficiently injecting the AO-corrected light into the chip in order to have a working photonic instrument for astronomy.

\subsection{Working principle of AWGs}

The main components inside the AWG chips in our laboratory are shown in detail in \autoref{fig:AWG_Emerald_Detailed}. Lets briefly follow the journey of light inside the AWG:
\begin{enumerate}
    \item A polychromatic beam of light is fed to the system through an input waveguide, coupled at its end to the input Free Propagation Region (FPR, also sometimes called a slab waveguide).
    \item As the beam enters the input FPR it is no longer laterally confined by the waveguide, so it starts diverging as it propagates towards the arrayed waveguides at the opposite end of the FPR. This entering beam is at the object plane of the system.
    \item The beam gets coupled into the array of single-mode waveguides closely spaced between each other. Efficient injection is achieved by the inclusion of coupling tapers that smoothly convert the super-mode that propagates through the FPR into the injected modes inside of the waveguides. This array has the appropriate geometry in order for each waveguide to be incrementally longer than the previous one by a constant difference $\Delta L$. At the end of the array of waveguides, each transported beam has a slight optical path (and thus phase) difference from that of its neighbor.
    \item The light from all the waveguides now gets coupled to an output FPR. The geometry of the FPR is such that at the output of the arrayed waveguides the coupled beams are converging towards the focusing surface of the FPR. The constant length increment in the waveguide array $\Delta L$ is tuned so that light at a central operation wavelength $\lambda_0$ will interfere constructively and get focused at the center of the output FPR.
    \item At the FPR's end surface, the reimaged beam gets focused onto its image plane. The phase differences between each of the focused beams generate a wavelength dependence on the location of their images, this effectively separates the different wavelengths along the image plane, and the dispersion is achieved.
    \item At this focal surface a spectrum is formed, satisfying the grating equation (adapted for the properties of the AWG). This means that different wavelengths, separated by the free spectral range (FSR) of the chip, will be imaged to the same position. This imposes an overlapping of the orders in this first spectrum that can be solved by cross-dispersion or spectral filtering after the output. In our chips this surface was discretely sampled using a fan-out of waveguides that separates each of the spectral channels and remaps them onto the flat output at the edge of the chip. This choice is particularly convenient since it turns the original curved focal surface into a flat one.
\end{enumerate}

The last step of remapping the image plane of the AWG into output waveguides is not a strict requisite. In some cases it could be desired to simply reimage the FPR's focal surface using bulk optics into a detector, or (if convenient) the detection can be performed directly on the focal surface. These alternatives can prove challenging due to the curved geometry of the FPR's focal plane.

\begin{figure}[t]
    \centering
    \includegraphics[width=1\linewidth]{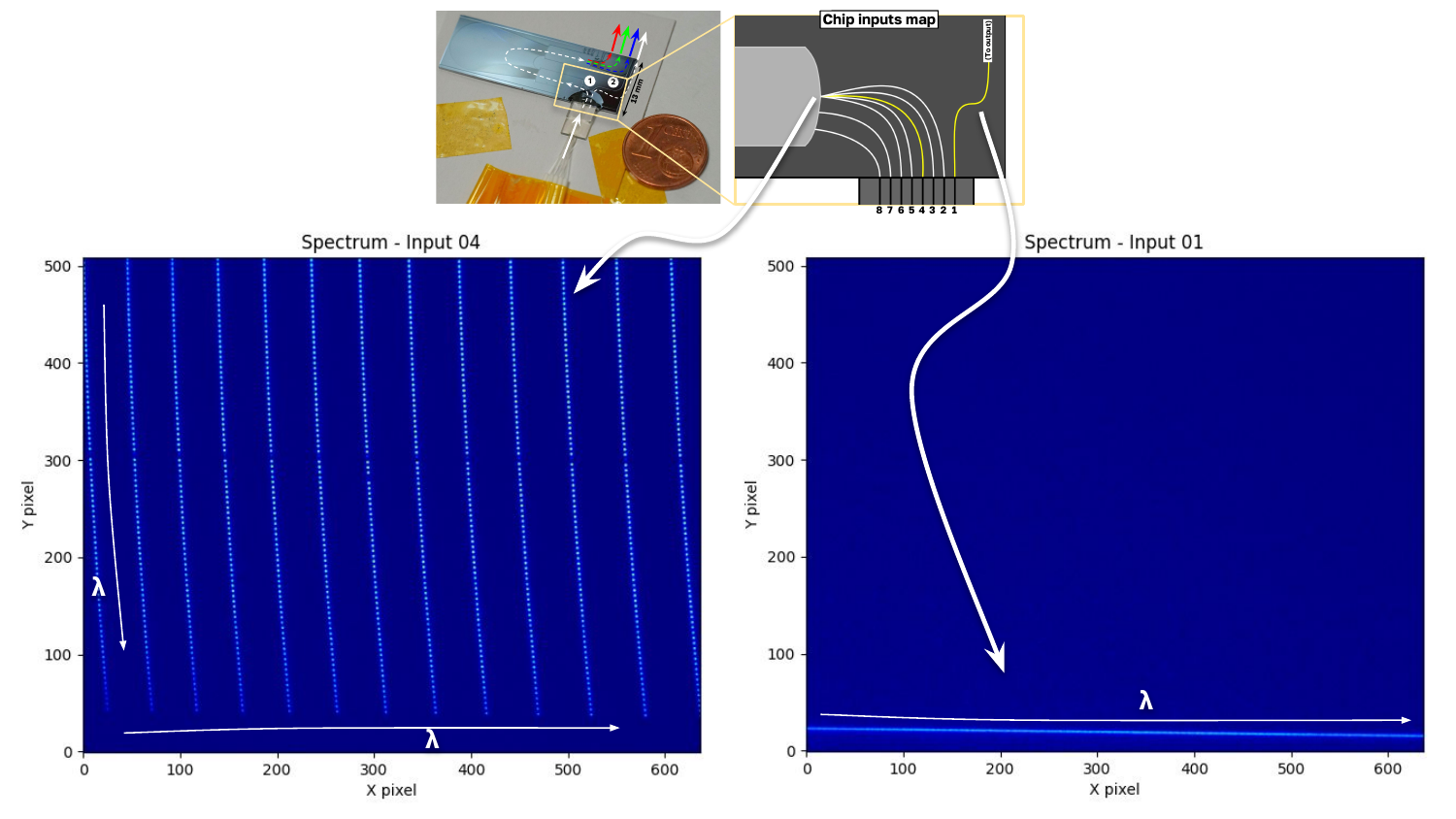}
    \caption{Comparison of spectra acquired by feeding light from a broadband halogen source through the AWG (injection into input 4, left) and through the bypassing or straight-through channel (injection into input 1, right). The direction of increasing wavelength along the orders and across the detector plane is shown. At the top a picture of the AWG Emerald chip with a 1 Euro cent coin is shown for reference. A zoomed in schematic of the inputs available in the case of our chips is shown.}
    \label{fig:Emerald_AWGvsVPH}
\end{figure}

In our case, after the output of the chip, a cross-dispersing back end (a simple setup of collimating lens - cross-disperser - focusing lens), separates the orders in the direction perpendicular to the first dispersion, and allows to reimage each spectral channel on a detector. \autoref{fig:Emerald_AWGvsVPH} illustrates how the spectra look like depending on the input used. It shows the result of injecting through an input feeding the AWG circuit (input 4) and injecting into the waveguide that bypasses the AWG and goes straight to the output (input 1, also called the straight-through waveguide). The specific setup for our prototype is shown in \autoref{fig:CoLiBRIS_Setup} and further explained in \autoref{sec:CoLiBRIS}.

The AWG spectrum shows the images of the cross-dispersed orders along the x axis, with wavelength increasing from left to right. Along each order the discrete spectral channels go increasing in wavelength from top to bottom. The dispersion along the vertical direction is thus due to the AWG and the along the horizontal direction is due to the cross-disperser (in this case a VPH grating). Changing the cross-disperser allows to change the spatial separation of the orders on the detector, which is important to balance wavelength coverage and crosstalk between orders. The cross-disperser (VPH) spectrum falls at the bottom of the detector plane, with wavelength increasing from left to right. The cross-disperser spectrum provides a lower resolution (and higher flux/pixel) reference for the AWG spectra.

\subsection{On-axis vs off-axis injection}

\begin{figure}[t]
    \centering
    \includegraphics[width=1\linewidth]{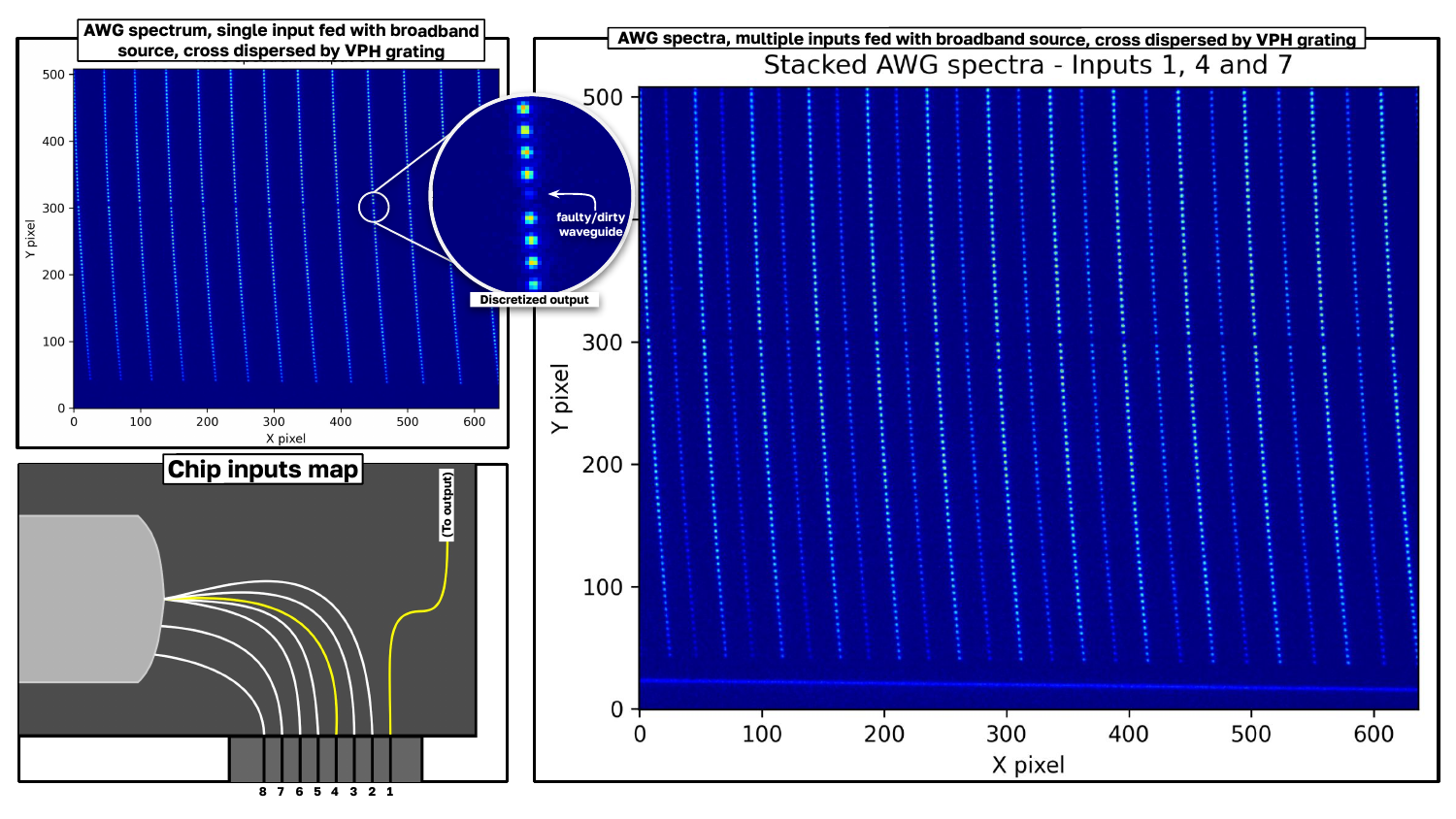}
    \caption{Overview of a typical AWG spectrum and of the different inputs available to feed the chip.}
    \label{fig:AWG_Emerald_inputs}
\end{figure}

In theory, the AWG's working principle holds as long as single-moded light is being injected close to the optical axis of the device (i.e. the center of the input FPR), but some useful applications can come out of injecting the light off axis. The effects of this can be seen in \autoref{fig:AWG_Emerald_inputs}, shown for the different inputs available in our chips.

If the dispersed output of the chip is then cross-dispersed and reimaged on the detector, feeding slightly off-axis light into the FPR leads to a spectrum with a horizontal offset on the detector plane. To understand how this happens it is enough to imagine a monochromatic beam being injected off-axis into the FPR: this offset at the object plane within the chip leads to an offset of the dispersed image at the focal plane of the second FPR, essentially causing the light to be injected into a different output waveguide than the one it would in the case of on-axis injection. After the light leaves the chip the cross-disperser will induce the horizontal displacement of the image of that channel to fall on a specific location on the detector. Since the light in this channel is of a different wavelength as in the on-axis case, this horizontal position will be different, resulting in the observed offset. 

In \autoref{fig:AWG_Emerald_inputs} this is shown in the right panel, where for the same field of view a case of on-axis injection (brighter spectrum) is stacked with an off-axis spectrum (slightly fainter spectrum). The VPH spectrum (from feeding the bypassing input 1) is shown in the bottom of the frame for reference.

\section{The CoLiBRIS AWG prototype}\label{sec:CoLiBRIS}

\begin{figure}[t]
    \centering
    \includegraphics[width=1\linewidth]{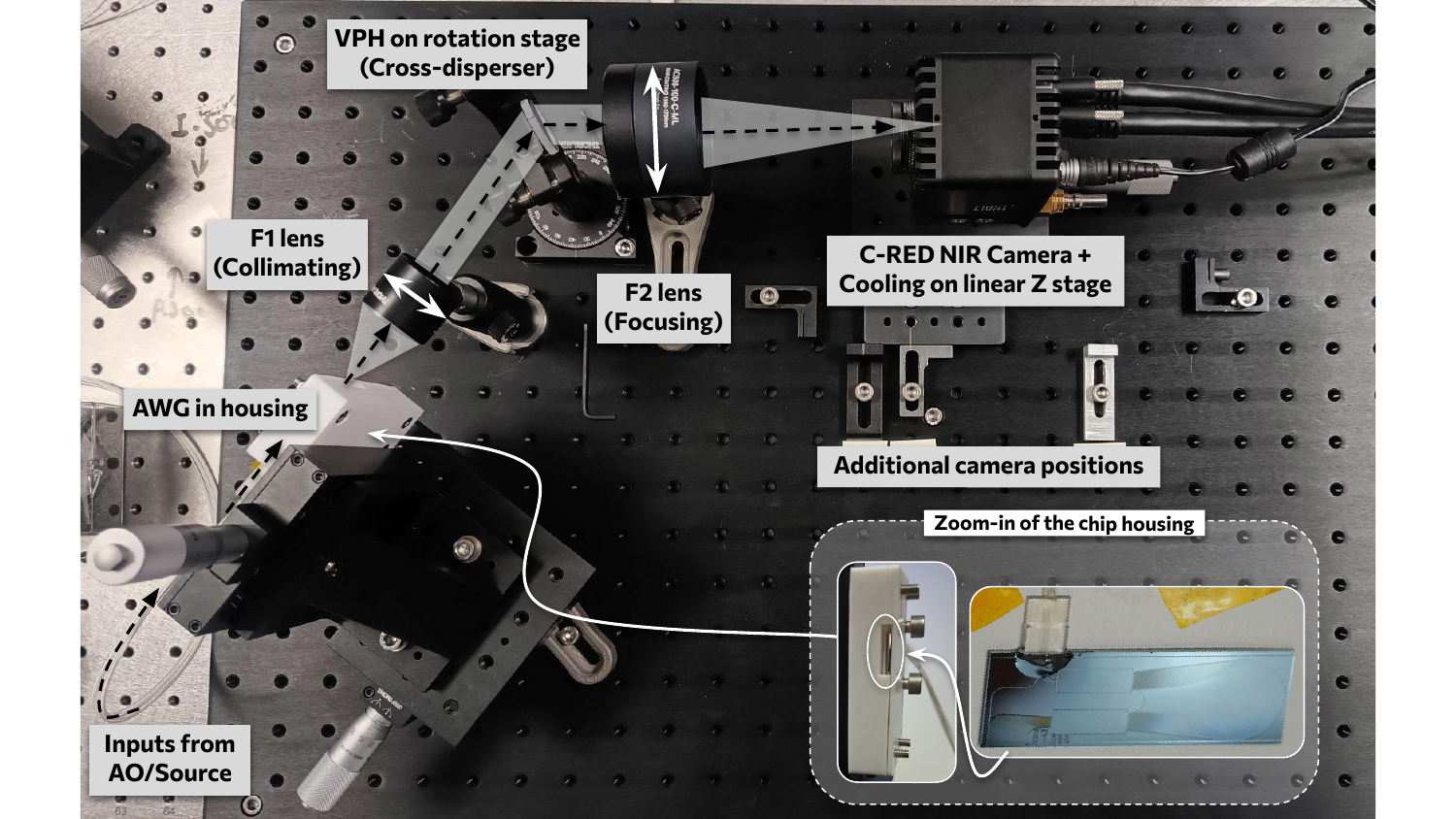}
    \caption{Schematic and bench setup of the CoLiBRIS spectrograph with the Emerald AWG chip mounted on a 3D-printed housing. The zoom-in at the bottom-right shows the chip inside of the 3D-printed housing with the output waveguides lit up with light from a broadband source.}
    \label{fig:CoLiBRIS_Setup}
\end{figure}

The \textbf{Compact Lightweight Broadband high-Resolution Infrared Spectrograph (CoLiBRIS-AWG)} prototype is a platform for testing AWG chips for astronomical spectroscopy. It was assembled for the characterization and testing of two AWG chips designed and manufactured by the team of N. Jovanovic and P. Gatkine in Caltech and UCLA, and later shipped to the Lagrange Laboratory of the Observatoire de la Côte d'Azur. The two SiN chips have been nicknamed \textit{Emerald} ($R\sim18800$, $13\times36 \text{ mm}^2$) and \textit{Sapphire} ($R\sim36000$, to be properly confirmed, $20\times66 \text{ mm}^2$). Both have an operational central wavelength of $1550 \text{ nm}$ and a free spectral range $\text{FSR}\sim 12 \text{ nm}$.

The setup of the CoLiBRIS spectrograph is shown in \autoref{fig:CoLiBRIS_Setup}. The whole instrument has the dimensions of $75\times45\times35 \text{ cm}^3$ with the bulk contribution in volume being the bulk-optics cross-dispersing backend that reimages the spectra onto the detector plane. It is composed by the 2-axis stage that holds the chip in its enclosure, a collimating lens (referred to as $f_1$), the VPH grating serving as a cross-disperser, a focusing lens (referred to as $f_2$), and a C-RED 3 infrared camera connected to the computer acquiring the images. The combination of $f_1$ and $f_2$ can be changed in order to test different degrees of magnification. 

We tested extensively the combination $f_1=40 \text{ mm}$ and $f_1=75 \text{ mm}$ which provided a good balance between channel sampling and broad wavelength coverage, while also allowing to keep the VPH spectrum of the bypassing straight-through waveguide in the bottom of the field of view of the camera. The downside of this choice is that around $30$ spectral channels at the shorter wavelength end of each order (the top of the image) do not fit in the field of view. To compensate for this the chip mount can be moved up and down with a linear stage, but datasets changing the height of the chip have not been explored yet.

The AWG spectra we acquire have a resemblance to conventional échelle spectra, with the particular feature of having discrete spectral channels, with a channel separation of $\sim0.08 \text{ nm}$ at $\lambda\sim 1560 \text{ nm}$. This configuration of the system provides a wavelength coverage of $1444 \text{ nm} < \lambda <1630 \text{ nm}$.

\section{Characterization and calibration}
\label{sec:sections}

With the CoLiBRIS prototype working in-lab we performed characterization and calibration measurements during the second half of 2025 to have a better knowledge of the chip and its capabilities. The Emerald chip has a total of 145 spectral channels per order of which only 2 were found to be faulty. The Sapphire chip has 290 channels per order of which 7 were not functioning properly. The zoomed in frame in \autoref{fig:AWG_Emerald_inputs} shows one of such faulty channels. These faulty waveguides can become a feature of the spectra that allows to have a point of reference to stich together spectra where the height of the chip can be varied to recover the missing channels at the top of the image.

With the help of super-continuum (Leukos Samba 450) and broadband halogen sources (Thorlabs SLS201L/M, resembling a black body spectrum) the alignment and focusing of the system was improved to decrease the effect of spherical aberrations from the lenses. This flat spectra were also used as a reference to characterize the response of the instrument and normalizing the spectra of other sources.

Using two tunable lasers (Santec TSL-510, and Thorlabs TLX1), we were able to record the spectra of individual monochromatic beams which were used to perform the spectral calibration of the system. The accuracy of the spectral solution needs to be further explored since the reference wavelengths provided by the tunable lasers do not cover the full wavelength range of the instrument.

\begin{figure}[t]
    \centering
    \includegraphics[width=1\linewidth]{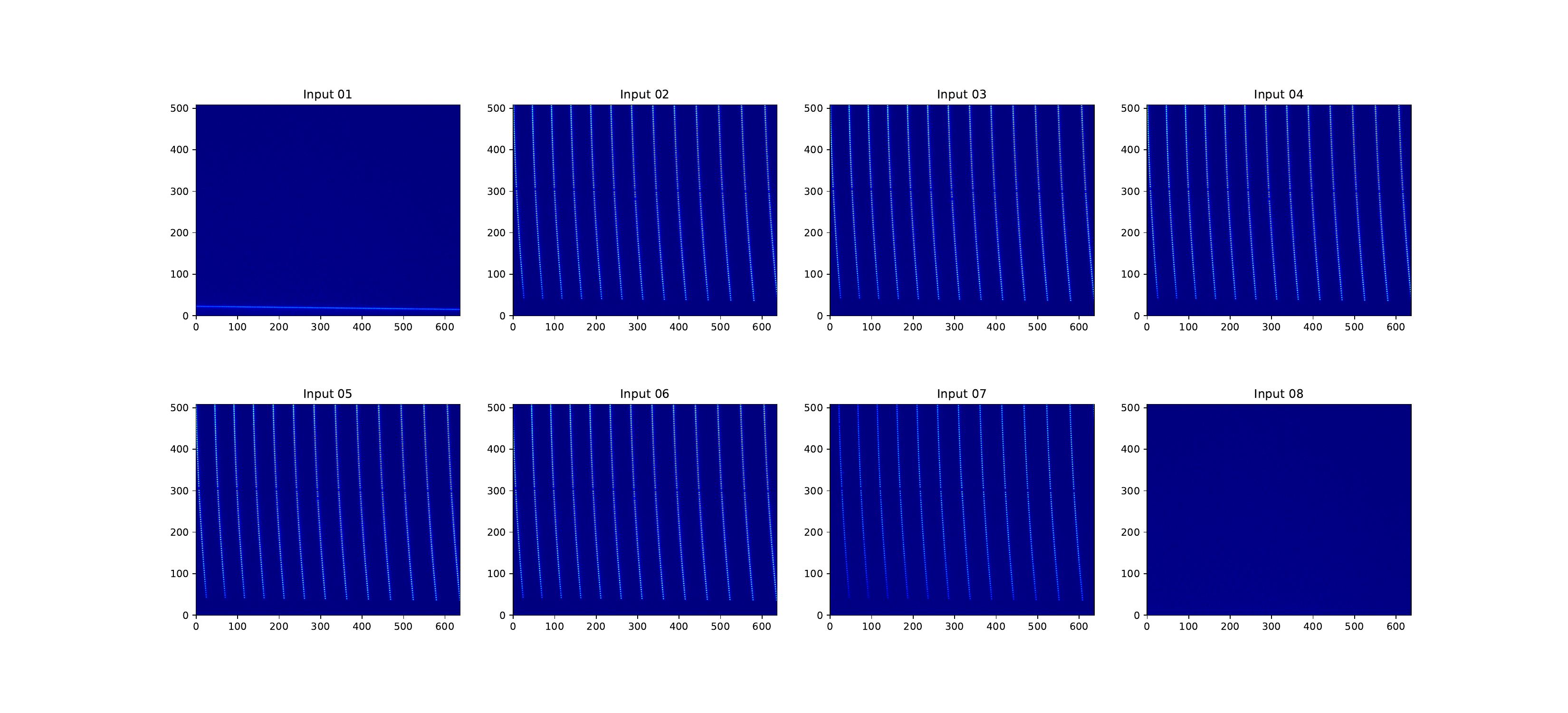}
    \caption{Spectra of the Thorlabs broadband source injected onto the 8 different inputs of the Emerald AWG chip.}
    \label{fig:AWG_Emerald_inputsgrid}
\end{figure}

Additionally, an initial exploration of feeding light through the different inputs of the Emerald chip was performed, leading to the spectra shown in \autoref{fig:AWG_Emerald_inputsgrid}. From this we know that the on-axis AWG inputs (2-6) seem to be functioning properly. Similarly the off-axis input (7) results in slightly fainter but usable spectra (as expected) offset from the on-axis ones. Input 8 which injects into the FPR from a highly off-axis position does not show any light going through the output of the chip.

\begin{figure}[t]
    \centering
    \includegraphics[width=1\linewidth]{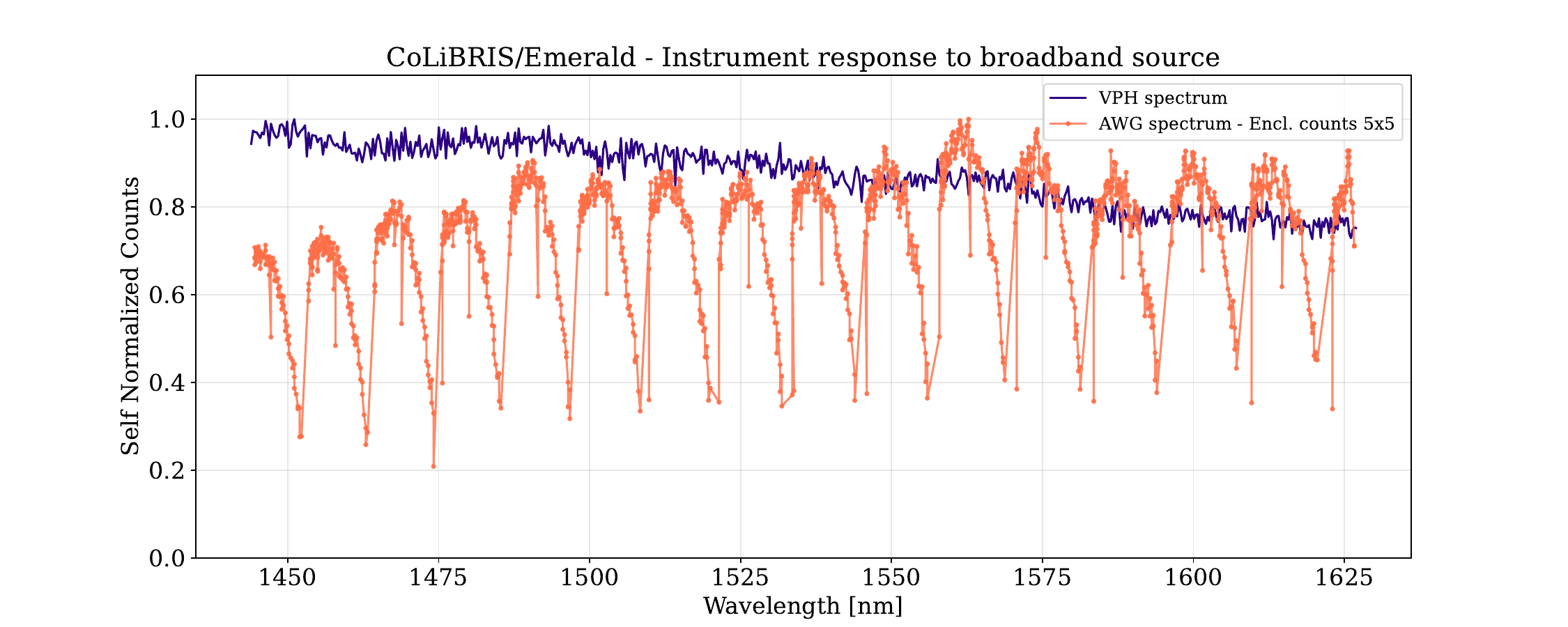}
    \caption{Instrument response to light from a Thorlabs SL201L broadband source. This source is used for calibration and for normalizing the flux of the stellar spectra. The solid blue line shows the low resolution VPH grating spectrum acquired by injecting into the straight-through input that allows bypassing the AWG circuit.}
    \label{fig:Spec_bb}
\end{figure}

Using the Thorlabs SLS201L/M broadband  source, the instrument's response to a ``flat" spectrum was recorded and is shown in \autoref{fig:Spec_bb}. The solid blue line shows the resulting spectrum from feeding the bypassing input of the chip, getting dispersion only from the VPH grating. The orange dots show the AWG, high-resolution spectrum. With these data and the spectra of the tunable laser spots it was measured that the total coverage of the spectrum was $1444 \text{ nm} < \lambda <1630 \text{ nm}$, and the wavelength separation between neighboring channels was $\sim0.08 \text{ nm}$ at $\lambda\sim 1560 \text{ nm}$. This measurement results in the estimation of the spectral resolution of $R\sim 18800$. The resolution of the spectra acquired only with the VPH grating is measured to be $R\sim 1500$.

An adequate estimation of the AWG's transmission (often called throughput) has not been performed. An initial assessment using enclosed counts around spectral channels using the tunable laser sources was made showing a consistent trend of relative transmission between the bypassing input and the AWG of $> 50 \%$. This figure is only an initial reference since many improvements on the background subtraction and proper flux estimation are imperative before providing an accurate estimate of the transmission. Additionally a characterization of channel cross-talk is necessary to understand better the real transmission of the instrument.

\section{On-sky testing}

We successfully performed first light observations in February 2026 at the focus of the T152 Telescope of the Observatory of Haute Provence (OHP) in France. AO correction was performed by the PAPYRUS system\cite{Fetick2023,Cisse2023}, that fed H-band starlight through a single-moded fiber into the CoLiBRIS prototype and the other spectrographs of the Compact Spectrographs collaboration (SWIFTS\cite{Martin2024} \& VIPA\cite{Carlotti2022}). 

The observing campaign focused on acquiring stellar spectra from bright K and M-type targets Arcturus ($\alpha$ Boo) and Betelgeuse ($\alpha$ Ori). Since these stars are bright ($H_{\text{mag}}<-2$) and have relevant absorption features in the H-band\cite{RaynerIRTF2009,CushingIRTF2005}. With the purposes of spectral normalization attempts on recording the spectra of A-type stars ($\zeta$ UMa and $\alpha$ Gem) were made, but their H-band flux was too low so these observations were dominated by noise. 

\begin{figure}[t]
    \centering
    \includegraphics[width=1\linewidth]{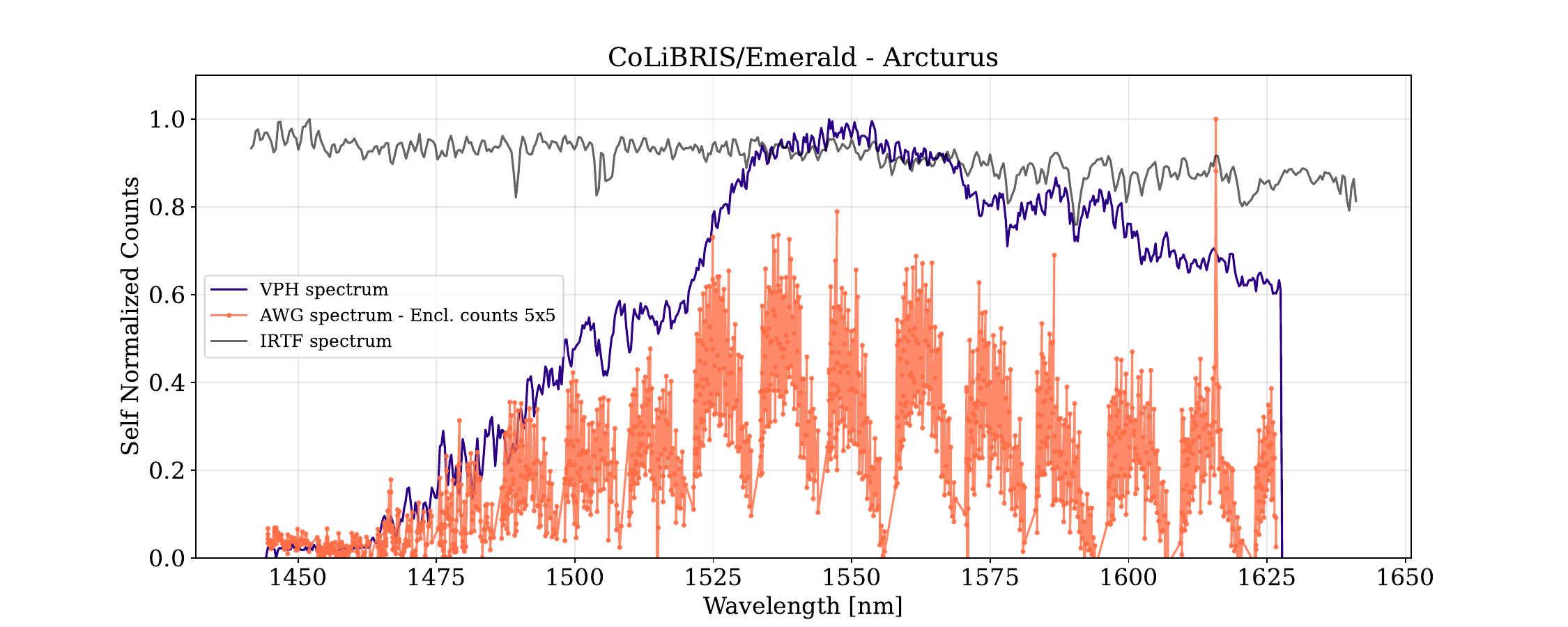}
    \caption{First light spectra of $\alpha$ Boo. The continuous gray line shows the reference catalog spectrum from the IRTF Spectral Library\cite{CushingIRTF2005,RaynerIRTF2009}. The continuous blue line shows the CoLiBRIS/Emerald low resolution VPH grating spectrum acquired by injecting into the straight-through input that allows bypassing the AWG circuit. The orange points show the preliminary extraction of the AWG spectrum.}
    \label{fig:Spec_Arcturus}
\end{figure}

\begin{figure}[t]
    \centering
    \includegraphics[width=1\linewidth]{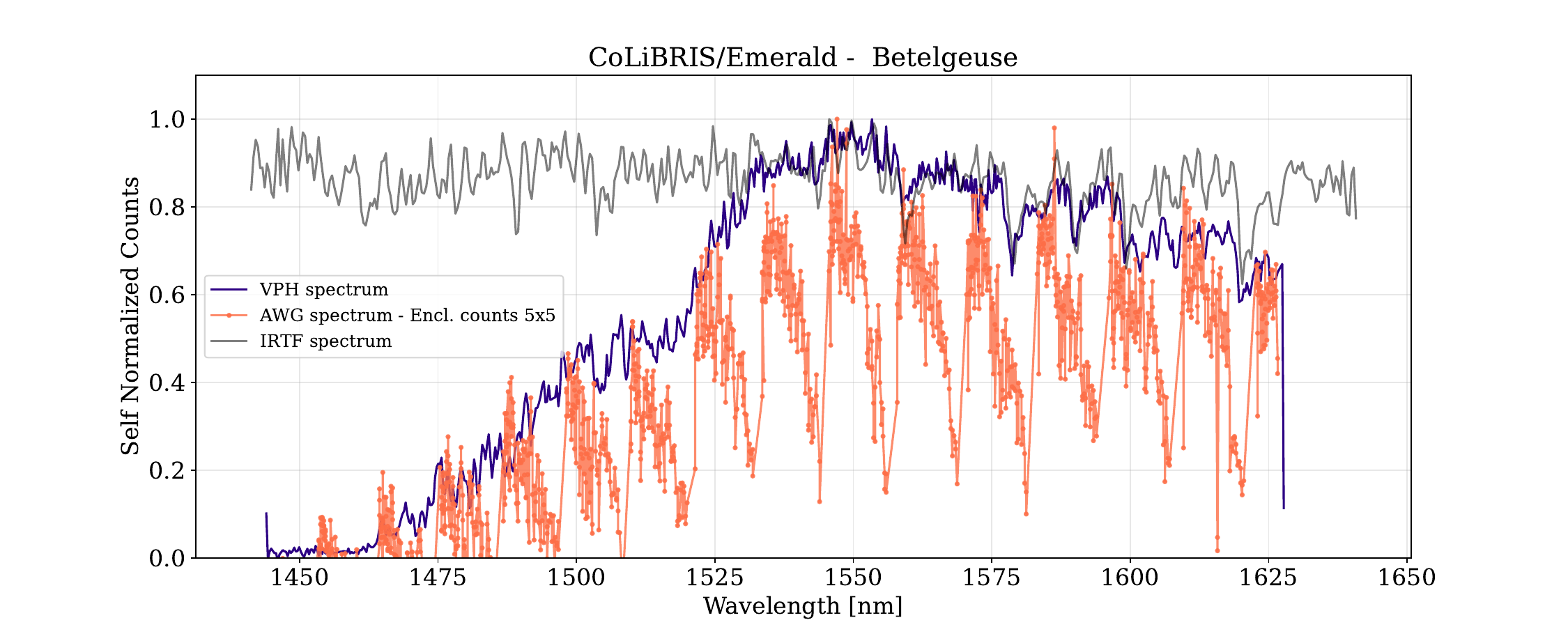}
    \caption{Same as \autoref{fig:Spec_Arcturus} but for $\alpha$ Ori.}
    \label{fig:Spec_Betelgeuse}
\end{figure}

Additional calibration images were taken using the tunable lasers and reference broadband sources that allowed us to calibrate the high-resolution stellar spectra acquired with the CoLiBRIS-AWG prototype. The first spectral extraction for two 20 s exposures of Arcturus and Betelgeuse are shown in \autoref{fig:Spec_Arcturus} and \autoref{fig:Spec_Betelgeuse}. 

These preliminary spectra show the AWG (from feeding starlight into input 4 of the chip), VPH (input 1), and a reference catalog spectrum for both targets. The reference spectra are from the IRTF Spectral Library\cite{CushingIRTF2005,RaynerIRTF2009}, acquired at a resolution of $R\sim 2000$ with the SpeX spectrograph at NASA's Infrared Telescope Facility (IRTF) in Mauna Kea, Hawai'i.

In both cases the absorption features in the VPH spectra match the location of the features in the IRTF spectra. Namely the Magnesium lines at $\sim 1580 \text{ nm}$ and the Iron doublet at $\sim 1505 \text{ nm}$. We see expected strong atmospheric absorption and telluric lines at the end of the J band ($\lambda < 1500 \text{ nm}$).

For the AWG spectra we see that even with flat normalization with the broadband source, the shape of the orders still has a strong wavelength dependence. It is also clear that better background subtraction and normalization techniques need to be implemented in order to clean up the AWG spectra. Nevertheless, the general shape of the spectra is consistent with what is seen in the VPH ones, and in the spectrum of Betelgeuse the shape of some absorption features are noticeably aligned with the absorption lines in the reference and VPH spectra.

\section{Conclusion and perspectives}

AWGs are promising alternatives for improving instrument footprint, stability and cost. Their resistance to environmental changes also makes them interesting candidates as space mission payloads. This first stage of characterization and on-sky testing shows that with only a few components a compact AWG-based astronomical spectrograph is feasible and functional. With our CoLiBRIS-AWG prototype we were able to acquire stellar spectra showing well known absorption features but that require a more robust extraction for adequate analysis. Nevertheless, this is to our knowledge the highest-resolution stellar spectra recorded with an AWG-based astronomical spectrograph reported to date. With a proper postprocessing pipeline (in development) we expect to deliver cleaner AWG spectra that can be properly analyzed in a following publication.

Further characterization and simulations of the system will follow. This includes a proper estimation of the transmission of the system (AWG + cross-dispersing backend), the effects of polarization, simultaneous injections through different inputs, and extending all these experiments to the Sapphire ($R\sim36000$) chip. This will help constrain better all the parameters and limitations of the instrument and will allow to perform an improved and more extensive observation campaign to apply the lessons learned so far. This will be both another opportunity to test the limits of our setup and also involve other AO-capable facilities in our project.

\begin{figure}[t]
    \centering
    \includegraphics[width=1\linewidth]{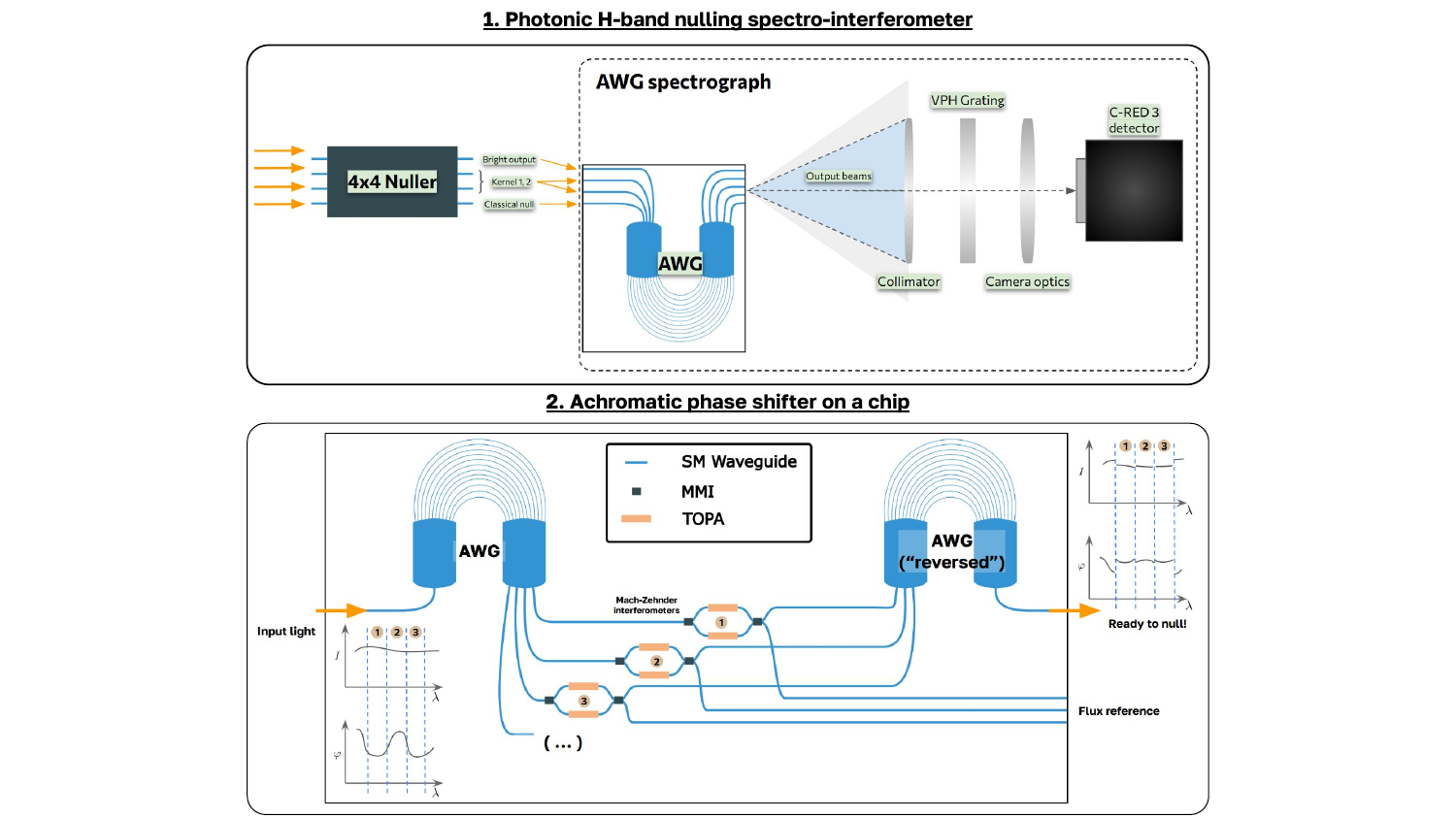}
    \caption{Two concepts being designed by our team to be implemented in the following months.}
    \label{fig:AWG_concepts}
\end{figure}

In parallel, two new concepts involving AWGs will be implemented and qualified in our lab over the following months. Schematics of these concepts are shown in \autoref{fig:AWG_concepts}. The first one is a simple setup dispersing the nulled output of a 4x4 kernel nulling chip. The spectrum of an off-axis source can be acquired by an AWG spectrograph at the null output. This would be a photonics-based spectro-interferometric prototype capable of achieving a deep null output and high spectral resolution. The second concept is the design of an achromatizing function of a nuller, akin to the so-called ``adaptive nuller", but $ 100\% $ photonics. Pairs of multiplexing - demultiplexing AWGs can be put on either end of on-chip complex-amplitude modulators to expand the bandwidth of the nuller by introducing phase shifts in specific spectral channels. This will be crucial to compensate the chromaticity of the components in the nuller.

\acknowledgments     
 
This work has benefited from state aid managed by the French \textbf{National Research Agency (ANR)} under the France 2030 program, bearing the reference \textbf{ANR-22-EXOR-0006}.

\noindent This work was supported by the \textbf{Action Spécifique Haute Résolution Angulaire (ASHRA) of CNRS/ INSU} co-funded by \textbf{CNES}.

\noindent This work was supported by the \textbf{Wilf Family Discovery Fund in Space and Planetary Science}, funded by the Wilf Family Foundation, as well as the support from Keck Institute for Space Studies at Caltech.

\bibliographystyle{spiebib}
\bibliography{references}

@Article{Jovanovic2023,
  author     = {Jovanovic, Nemanja and Gatkine, Pradip and Anugu, Narsireddy and Amezcua-Correa, Rodrigo and Basu Thakur, Ritoban and Beichman, Charles and Bender, Chad F. and Berger, Jean-Philippe and Bigioli, Azzurra and Bland-Hawthorn, Joss and Bourdarot, Guillaume and Bradford, Charles M and Broeke, Ronald and Bryant, Julia and Bundy, Kevin and Cheriton, Ross and Cvetojevic, Nick and Diab, Momen and Diddams, Scott A and Dinkelaker, Aline N and Duis, Jeroen and Eikenberry, Stephen and Ellis, Simon and Endo, Akira and Figer, Donald F and Fitzgerald, Michael P. and Gris-Sanchez, Itandehui and Gross, Simon and Grossard, Ludovic and Guyon, Olivier and Haffert, Sebastiaan Y and Halverson, Samuel and Harris, Robert J and He, Jinping and Herr, Tobias and Hottinger, Philipp and Huby, Elsa and Ireland, Michael and Jenson-Clem, Rebecca and Jewell, Jeffrey and Jocou, Laurent and Kraus, Stefan and Labadie, Lucas and Lacour, Sylvestre and Laugier, Romain and Ławniczuk, Katarzyna and Lin, Jonathan and Leifer, Stephanie and Leon-Saval, Sergio and Martin, Guillermo and Martinache, Frantz and Martinod, Marc-Antoine and Mazin, Benjamin A and Minardi, Stefano and Monnier, John D and Moreira, Reinan and Mourard, Denis and Nayak, Abani Shankar and Norris, Barnaby and Obrzud, Ewelina and Perraut, Karine and Reynaud, François and Sallum, Steph and Schiminovich, David and Schwab, Christian and Serbayn, Eugene and Soliman, Sherif and Stoll, Andreas and Tang, Liang and Tuthill, Peter and Vahala, Kerry and Vasisht, Gautam and Veilleux, Sylvain and Walter, Alexander B and Wollack, Edward J and Xin, Yinzi and Yang, Zongyin and Yerolatsitis, Stephanos and Zhang, Yang and Zou, Chang-Ling},
  journal    = {Journal of Physics: Photonics},
  title      = {2023 Astrophotonics Roadmap: pathways to realizing multi-functional integrated astrophotonic instruments},
  year       = {2023},
  issn       = {2515-7647},
  month      = oct,
  number     = {4},
  pages      = {042501},
  volume     = {5},
  doi        = {10.1088/2515-7647/ace869},
  publisher  = {IOP Publishing},
  readstatus = {read},
}

@InBook{Leijtens2006,
  author    = {Leijtens, Xaveer J. M. and Kuhlow, Berndt and Smit, Meint K.},
  pages     = {125--187},
  publisher = {Springer Berlin Heidelberg},
  title     = {Arrayed Waveguide Gratings},
  year      = {2006},
  isbn      = {9783540317692},
  booktitle = {Wavelength Filters in Fibre Optics},
  doi       = {10.1007/3-540-31770-8_5},
}

@Article{Smit1988,
  author    = {Smit, M.K.},
  journal   = {Electronics Letters},
  title     = {New focusing and dispersive planar component based on an optical phased array},
  year      = {1988},
  issn      = {1350-911X},
  month     = mar,
  number    = {7},
  pages     = {385--386},
  volume    = {24},
  doi       = {10.1049/el:19880260},
  publisher = {Institution of Engineering and Technology (IET)},
}

@Article{Cvetojevic2012,
  author    = {Cvetojevic, N. and Jovanovic, N. and Betters, C. and Lawrence, J. S. and Ellis, S. C. and Robertson, G. and Bland-Hawthorn, J.},
  journal   = {Astronomy \& Astrophysics},
  title     = {First starlight spectrum captured using an integrated photonic micro-spectrograph},
  year      = {2012},
  issn      = {1432-0746},
  month     = jul,
  pages     = {L1},
  volume    = {544},
  doi       = {10.1051/0004-6361/201219116},
  publisher = {EDP Sciences},
}

@Article{Takahashi1990,
  author    = {Takahashi, H. and Suzuki, S. and Kato, K. and Nishi, I.},
  journal   = {Electronics Letters},
  title     = {Arrayed-waveguide grating for wavelength division multi/demultiplexer with nanometre resolution},
  year      = {1990},
  issn      = {1350-911X},
  month     = jan,
  number    = {2},
  pages     = {87--88},
  volume    = {26},
  doi       = {10.1049/el:19900058},
  publisher = {Institution of Engineering and Technology (IET)},
}

@Article{Fetick2023,
  author    = {Fetick, Romain and Chambouleyron, Vincent and Muslimov, Edouard and Boudjema, Idir and Striffling, Arnaud and Cisse, Mahawa and Heritier, Cedric Taissir and Pedreros Bustos, Felipe and Soria HernAndez, Esther and Taylor, Jacob and Leroux, Francois and Camelo, Raissa and Jouve, Pierre and Levraud, Nicolas and Schmitt, Jerome and Sauvage, Jean-Francois and Martin, Bruno and El-Hadi, Jacem and Charton, Julirn and Neichel, Benoit and Fusco, Thierry},
  title     = {PAPYRUS: one year of on-sky operations},
  year      = {2023},
  copyright = {CC-BY},
  doi       = {10.13009/AO4ELT7-2023-014},
  language  = {Eng},
  publisher = {Proceedings of AO4ELT7 conference 2023 - Editors: Thierry Fusco and Benoit Neichel},
}

@ARTICLE{CushingIRTF2005,
       author = {{Cushing}, Michael C. and {Rayner}, John T. and {Vacca}, William D.},
        title = "{An Infrared Spectroscopic Sequence of M, L, and T Dwarfs}",
      journal = {\apj},
         year = 2005,
        month = apr,
       volume = {623},
       number = {2},
        pages = {1115-1140},
          doi = {10.1086/428040},
archivePrefix = {arXiv},
       eprint = {astro-ph/0412313},
 primaryClass = {astro-ph},
       adsurl = {https://ui.adsabs.harvard.edu/abs/2005ApJ...623.1115C}
}

@ARTICLE{RaynerIRTF2009,
       author = {{Rayner}, John T. and {Cushing}, Michael C. and {Vacca}, William D.},
        title = "{The Infrared Telescope Facility (IRTF) Spectral Library: Cool Stars}",
      journal = {\apjs},
         year = 2009,
        month = dec,
       volume = {185},
       number = {2},
        pages = {289-432},
          doi = {10.1088/0067-0049/185/2/289},
archivePrefix = {arXiv},
       eprint = {0909.0818},
 primaryClass = {astro-ph.SR},
       adsurl = {https://ui.adsabs.harvard.edu/abs/2009ApJS..185..289R}
}

@INPROCEEDINGS{Carlotti2022,
       author = {{Carlotti}, Alexis and {Bidot}, Alexis and {Mouillet}, David and {Correia}, Jean-Jacques and {Jocou}, Laurent and {Curaba}, Stephane and {Delboulb{\'e}}, Alain and {Le Coarer}, Etienne and {Rabou}, Patrick and {Bourdarot}, Guillaume and {Forveille}, Thierry and {Bonfils}, Xavier and {Vasisht}, Gautam and {Mawet}, Dimitri and {Burruss}, Rick S. and {Oppenheimer}, Rebecca and {Doyon}, Ren{\'e} and {Artigau}, Etienne and {Vall{\'e}e}, Philippe},
        title = "{On-sky demonstration at Palomar Observatory of the near-IR, high-resolution VIPA spectrometer}",
    booktitle = {Ground-based and Airborne Instrumentation for Astronomy IX},
         year = 2022,
       editor = {{Evans}, Christopher J. and {Bryant}, Julia J. and {Motohara}, Kentaro},
       series = {Society of Photo-Optical Instrumentation Engineers (SPIE) Conference Series},
       volume = {12184},
        month = aug,
          eid = {121841I},
        pages = {121841I},
          doi = {10.1117/12.2628937},
archivePrefix = {arXiv},
       eprint = {2305.19736},
 primaryClass = {astro-ph.IM},
       adsurl = {https://ui.adsabs.harvard.edu/abs/2022SPIE12184E..1IC}
}

@INPROCEEDINGS{Martin2024,
       author = {{Martin}, Guillermo and {Baccar}, Salma and {Mestre}, No{\'e}mie and {Bonduelle}, Myriam and {robert}, laurent and {Salut}, Roland and {Courjal}, Nadege and {Morand}, Alain},
        title = "{Development of fundamental buildings blocks needed for high spectral range integrated optics spectrometry: active phase modulation to increase sampling efficiency in Fourier transform spectrometers}",
    booktitle = {Advances in Optical and Mechanical Technologies for Telescopes and Instrumentation VI},
         year = 2024,
       editor = {{Navarro}, Ram{\'o}n and {Jedamzik}, Ralf},
       series = {Society of Photo-Optical Instrumentation Engineers (SPIE) Conference Series},
       volume = {13100},
        month = aug,
          eid = {131001X},
        pages = {131001X},
          doi = {10.1117/12.3018928},
       adsurl = {https://ui.adsabs.harvard.edu/abs/2024SPIE13100E..1XM}
}

@INPROCEEDINGS{Cisse2023,
       author = {{Cisse}, Mahawa and {Muslimov}, Eduard and {Taissir Heritier}, Cedric and {Chambouleyron}, Vincent and {Fetick}, Romain and {Levraud}, Nicolas and {Sauvage}, Jean-Fran{\c{c}}ois and {Neichel}, Benoit and {Fusco}, Thierry},
        title = "{PAPYRUS: Second stage adaptive optics with a vector Zernike wavefront sensor}",
    booktitle = {Adaptive Optics for Extremely Large Telescopes (AO4ELT7)},
         year = 2023,
        month = jun,
          eid = {58},
        pages = {58},
          doi = {10.13009/AO4ELT7-2023-060},
       adsurl = {https://ui.adsabs.harvard.edu/abs/2023aoel.confE..58C}
}

@ARTICLE{Gatkine2017,
       author = {{Gatkine}, Pradip and {Veilleux}, Sylvain and {Hu}, Yiwen and {Bland-Hawthorn}, Joss and {Dagenais}, Mario},
        title = "{Arrayed waveguide grating spectrometers for astronomical applications: new results}",
      journal = {Optics Express},
         year = 2017,
        month = jul,
       volume = {25},
       number = {15},
        pages = {17918},
          doi = {10.1364/OE.25.017918},
archivePrefix = {arXiv},
       eprint = {1707.03445},
 primaryClass = {astro-ph.IM},
       adsurl = {https://ui.adsabs.harvard.edu/abs/2017OExpr..2517918G}
}

@ARTICLE{Gatkine2021,
       author = {{Gatkine}, Pradip and {Jovanovic}, Nemanja and {Hopgood}, Christopher and {Ellis}, Simon and {Broeke}, Ronald and {{\L}awniczuk}, Katarzyna and {Jewell}, Jeffrey and {Wallace}, J. Kent and {Mawet}, Dimitri},
        title = "{Potential of commercial SiN MPW platforms for developing mid/high-resolution integrated photonic spectrographs for astronomy}",
      journal = {\ao},
         year = 2021,
        month = jul,
       volume = {60},
       number = {19},
        pages = {D15},
          doi = {10.1364/AO.423439},
archivePrefix = {arXiv},
       eprint = {2106.04598},
 primaryClass = {astro-ph.IM},
       adsurl = {https://ui.adsabs.harvard.edu/abs/2021ApOpt..60D..15G}
}

\end{document}